\documentclass[11pt]{article}
\usepackage{times, amsthm,amsfonts, amsmath, amssymb, latexsym, setspace}
\usepackage{endfloat}

\notablist

\def\bs{\boldsymbol}
\def\Ex{{\rm I\!E}}
\def\Pr{{\rm I\!P}}
\def\be{\begin{equation}}
\def\ee{\end{equation}}
\def\bea{\begin{eqnarray*}}
\def\eea{\end{eqnarray*}}
\def\bean{\begin{eqnarray}}
\def\eean{\end{eqnarray}}
\def\nn{\nonumber}
\def\nin{\noindent}

\def\ra{\rightarrow}

\def\Bl{\Bigl}
\def\Br{\Bigr}

\def\R{{\bs{R}}}
\def\II{{\cal I}}
\def\JJ{{\cal J}}

\def\eps{\epsilon}
\def\lam{\lambda}

\def\Iapp{\bs{\II}_{app}}
\def\Japp{\bs{\JJ}_{app}}

\def\lLR{logLR_n}

\newtheorem{Theorem}{Theorem}
\newtheorem{Proposition}{Proposition}

\newtheorem{Lemma}{Lemma}

\begin{document}
\setstretch{1.5}

\title{Optimal detection of a jump in the intensity of a Poisson process or 
in a density with likelihood ratio statistics}
\author{Camilo Rivera and Guenther Walther${}^{*}$ \\
        Stanford University}
\date{revised April 2013}
\maketitle

\begin{abstract}
We consider the problem of detecting a `bump' in the intensity of a Poisson 
process or in a density.
We analyze two types of likelihood ratio based statistics
which allow for exact finite 
sample inference and asymptotically optimal detection:
The maximum of the penalized square root of log likelihood ratios (`penalized
scan') evaluated over a certain sparse set of intervals, and a certain
average of log likelihood ratios (`condensed average likelihood ratio').
We show that penalizing the {\sl square root} of the log likelihood ratio 
- rather than the log likelihood ratio itself -
leads to a simple penalty term that yields optimal power.
The thus derived penalty may prove useful for other problems that involve
a Brownian bridge in the limit. The second key tool is an
approximating set of intervals that is rich enough to allow for optimal
detection but which is also sparse
enough to allow justifying the validity of the penalization scheme
simply via the union bound. This results in a considerable simplification
in the theoretical treatment compared to the usual approach for
this type of penalization technique, which
requires establishing an exponential inequality for the variation of the
test statistic. Another advantage of using the sparse approximating set is
that it allows fast computation in nearly linear time.

We present a simulation study that illustrates the superior performance
of the penalized scan and of the condensed average likelihood ratio compared to
the standard scan statistic.
\end{abstract}

\vfill


\noindent\textbf{Running headline}: Jump detection with likelihood ratios.


\noindent${}^{*}$ Work supported by NSF grant DMS-1007722 and NIH grant
AI09851901

\newpage

\section{Introduction and overview of results} \label{intro}

The paper is concerned with the following problem: One observes
an inhomogeneous Poisson process $X_1,\ldots,X_N$ on the real
line with intensity
$$
\lam (x)=
\begin{cases}
p \mu (x),& x \in I\\
q \mu (x),& x \not\in I
\end{cases}
$$
where $\mu(x) \geq 0$ is a known function with $\int \mu < \infty$,
but $p,q > 0$ and the interval $I$ are unknown. Hence the intensity
is known up to a multiplicative factor and we want to test whether
this factor is elevated on some interval $I$:
$$
H_0:\ p=q,\ \ \ H_A: p>q\ \mbox{ for some interval $I$.}
$$
This setting arises in a number of applications involving the detection
of a `cluster', see e.g. Glaz and Balakrishnan~(1999),
Loader~(1991) and Kulldorff~(1997). The latter two references
also give extensions to the bivariate
case, which is relevant for detecting spatial disease clusters 
while adjusting for the known population density $\mu$.
Since under $H_0$ the nuisance parameter $p=q$ is unknown,
we follow Loader~(1991) and analyze the problem
conditional on $N=n$. Then $X_1,\ldots,X_n$ are i.i.d. with density
\be \label{density}
f_{r,I}(x)=\frac{r1(x\in I)+1(x \in I^c)}{rF_0(I)+F_0(I^c)} f_0(x),\ \ 
\mbox{ where } f_0(x):=\frac{\mu(x)}{\int \mu}\ \mbox{ and }r:=\frac{p}{q},
\ee
and the testing problem becomes $H_0:\, r=1$ vs. $H_A:\, r>1$, so we test
whether the observations come from a known density $f_0$ (which we may
assume w.l.o.g. to be the uniform density, see (\ref{density2})) vs. the case
where $f_0$ is elevated by a multiplicative factor over some interval $I$.
Thus the methodology introduced in this paper may also be applied
for certain `bump-hunting' problems, see e.g. Good and Gaskins~(1980),
Hartigan~(1985), M\"{u}ller and Sawitzki~(1991), Minnotte and Scott~(1993)
 or Polonik~(1995). 

Loader~(1991) and Kulldorff~(1997) address the above problem with
the scan statistic, i.e. the maximum of the log likelihood ratio
statistic for varying $I$.
Chan and Walther~(2013) investigate a related problem in
the abstract Gaussian White Noise model. They show that the
scan is generally suboptimal for this type of detection problem,
but that optimal detection is possible by averaging
likelihood ratios over a judiciously chosen collection
of intervals. They also suggest that optimality can be restored for
the scan either by modifying it with a penalty term
that was introduced by D\"{u}mbgen and Spokoiny~(2001) for kernel
statistics in a different context, or by using the blocked scan
introduced by Walther~(2010) and Rufibach and Walther~(2010).

Here we show how optimal detection can be achieved in the practically
important case of intensities and densities with
likelihood ratios as the principal tool for inference. The main problem 
in trying to
adapt the penalization technique from the abstract
Gaussian White Noise model is that the form of the penalty
term depends partly on the specifics of an exponential inequality
that needs to be established for the variation of the local
test statistic. This inequality has to be established anew in each
setting, and this is a quite difficult theoretical exercise, see
Section~\ref{Pnallproof}.
Walther~(2010) and Rufibach and Walther~(2010) circumvent this
problem by penalizing p-values rather than critical values,
but at the cost of a more complex methodology and more computation.

One of the main contributions of this paper is to show how the
conceptually simpler penalization of critical values can be
implemented in the important case of log likelihood ratios, without
having to establish an exponential inequality for its variation.
Our main tool is to consider an appropriate subcollection of the
collection of all intervals. It is possible to construct such an approximating
set of intervals that on the one hand is rich
enough to allow optimal detection and on the other hand is sparse
enough to allow justifying the validity of the penalization scheme 
simply with the union bound. This approach was used in Walther~(2010)
in the multivariate Bernoulli model to penalize p-values when
scanning with rectangles. Our key idea to make this approach work
for penalizing critical values is to penalize the square root of twice 
the log likelihood
ratio instead of the log likelihood ratio. This transformation results
in a penalty that yields optimal detection. And due to the use of a sparse
approximating set of intervals, the appropriate
penalty term can be read off from the tail bound of the log likelihood
ratio itself, which in this case is simply given by Hoeffding's inequality.
As will become
clear from the exposition, this methodology should also be applicable
in a wide range of other contexts, such as those cited in this section.

We end up with a new penalty that
is somewhat different from the one used in D\"{u}mbgen and Spokoiny~(2001).
The form of this new penalty derives from a different limiting process (Brownian
bridge instead of Brownian motion) and simulations show that it results
in a superior finite sample performance when compared to
the D\"{u}mbgen-Spokoiny penalty. 

In the second part of the paper we show that
averaging the likelihood ratios over a particular approximating 
set of intervals (the {\it condensed average likelihood ratio (ALR)})
also results in optimal detection. We note that the construction of an
appropriate approximating set of intervals plays a crucial role for
both methodologies, both in terms of statistical inference
and for efficient computation: For the condensed ALR, the appropriate
construction of an approximating set directly results in optimal detection, 
while for the penalized scan
it justifies the use of the particular penalty term. In both cases
it results in efficient algorithms that run in almost linear time
versus the quadratic algorithms required for evaluating all intervals.
This computational aspect may well be the dominant concern for some users.

In Section~\ref{simulations} we provide a simulation
study that shows that the penalized scan and the condensed ALR
are clearly superior to the scan, with the condensed ALR having the overall
best performance.

\section{The scan statistic and the penalized scan}
\label{scan}

We will work in the density setting (\ref{density}), i.e. conditional
on $N=n$. The main advantage of such a conditional analysis is that
it eliminates the nuisance parameter $p$ under the null hypothesis,
and hence this approach avoids the problematic performance of likelihood 
ratio tests when a parameter is misspecified. Another advantage of the
conditional analysis is that it allows for exact finite sample inference
as will be seen below. Finally, we note that the conditional analysis does
not require the underlying point process to be a Poisson process, but it
is also valid for certain other processes that are not Poisson processes or
that do not even have independent increments.

A standard computation shows that for a given interval $I$ the log likelihood
ratio test statistic for testing $H_0:\, r=1$ vs. $H_A:\, r>1$ in 
(\ref{density}) is given by
$$
\lLR (F_0(I),F_n(I))\ :=
\begin{cases}
 n F_n(I) \log\Bl( \frac{F_n(I)}{F_0(I)}\Br) +n(1-F_n(I))
\log\Bl( \frac{1-F_n(I)}{1-F_0(I)}\Br)\;\;\; \mbox{ if $F_n(I)>F_0(I)$}\\
0\;\;\; \mbox{ else,}
\end{cases}
$$
where $F_n$ denotes the empirical cdf. Since $I$ is unknown, it is
customary to assess the evidence against $H_0$ with the {\sl scan
statistic} (maximum likelihood ratio statistic)
\begin{align}
M_n\,&:=\, \sup_{\mbox{intervals } I \subset \R} \lLR \Bl(F_0(I),F_n(I)\Br) 
     \nn \\
     &= \max_{1 \leq j<k \leq n} \lLR 
 \Bl(F_0([X_{(j)},X_{(k)}]),\frac{k-j+1}{n}\Br) \label{Mn}
\end{align}
where the equality follows from elementary considerations. Kulldorff~(1997)
gives a derivation of the maximum likelihood ratio without conditioning
on $N$ that results in the same formula for $M_n$. As observed experimentally 
by Neill~(2009a) and Chan~(2009) and explained theoretically by
Chan and Walther~(2013) in an abstract Gaussian regression setting, the scan 
will generally be suboptimal for detection. One way to rectify the situation 
is by adding a penalty term as introduced by D\"{u}mbgen and Spokoiny~(2001)
for kernel estimates. We propose to use the following form for a
{\sl penalized scan}:
$$
P_n\,:=\, \max_{[X_{(j)},X_{(k)}] \in \Japp}
\Bl(\sqrt{2 \lLR\Bl(F_0([X_{(j)},X_{(k)}]),\frac{k-j+1}{n}\Br)}
-\sqrt{2 \log \frac{en^2}{(k-j)(n-k+j)}}\Br),
$$
where the data-dependent collection of intervals $\Japp$ is defined below. 
For some applications it may be more appropriate to use a collection of
intervals that is not data-dependent, see e.g. Neill~(2009b). We therefore
also analyze the variant
$$
P_n^0\,:=\, \max_{I \in \Japp^0}
\Bl(\sqrt{2 \lLR\Bl(F_0(I),F_n(I)\Br)}
-\sqrt{2 \log \frac{e}{F_n(I)\Bl(1-\min(F_n(I),\frac{1}{2})\Br)}}\Br),
$$
where $\Japp^0$ is defined below. Note that the structure of the penalty 
in $P_n^0$ is
essentially the same as that in $P_n$, but a different notation is
required since the intervals in $\Japp^0$ are not
determined by the data. The null distributions of both $P_n$ and $P_n^0$
are distribution free, which allows exact finite sample inference
as detailed in Section~\ref{simulations}. Penalizing
the square root of $\lLR$ instead of $\lLR$ is crucial if one wants
to use a simple, additive penalty term that yields optimal detection:
Calculations show that an analogously derived  penalty term for $\lLR$
will not result in
optimal detection, unless one is willing to work with an intricate
non-additive penalty. The above penalty is different from
what one would expect from the work in the abstract Gaussian settings
in D\"{u}mbgen and Spokoiny~(2001) and Chan and Walther~(2013). That work
suggests to penalize the statistic pertaining to the interval $I$ with 
$\sqrt{2 \log e/F_n(I)}$. However, it will be seen in Section~\ref{Pnallproof}
 that the
relevant limiting process of $\sqrt{\lLR}$ does not involve the increments
of Brownian motion but those of the Brownian bridge. While a theoretical 
analysis
shows that one can still employ the $\sqrt{2 \log e/F_n(I)}$ penalty
for the latter case (provided that $F_n(I)$ stays bounded away from 1),
it also shows that there is some flexibility in designing the penalty. 
In fact, the theoretical analysis in Section~\ref{Pnallproof}  as well as   
simulations show that for a Brownian bridge 
it is much preferable to use the penalty
$\sqrt{2 \log \frac{e}{F_n(I)(1-\min(F_n(I),\frac{1}{2}))}}$, 
and this is essentially the
penalty we used for $P_n$ since we always have $F_n(I)\leq \frac{1}{2}$
there.

As approximating set $\Japp$ we use the univariate version of the
approximating set introduced in Walther~(2010):

\begin{align*}
\Japp &= \bigcup_{\ell =2}^{\ell_{max}} \Japp(\ell),\;\;\; \mbox{ where }
\;\ell_{max}=\Bl\lfloor \log_2 \frac{n}{\log n}\Br\rfloor \;\; 
\mbox{ and}\\
\Japp(\ell) &= \Bl\{[X_{(j)},X_{(k)}]:
j,k \in \{1+i d_{\ell},
i=0,1,\ldots \}\ \mbox{ and } m_{\ell}<k-j\leq 2m_{\ell}\Br\},\\
&\;\;\;\;\mbox{ where } m_{\ell}=n2^{-\ell},\; d_{\ell}=
\Bl\lceil \frac{m_{\ell}}{6 \sqrt{\ell}}\Br\rceil.
\end{align*}

$\Japp^0$ is defined\footnote{$\log_2$ and $\log$ denote the logarithm 
with base 2 and $e$, respectively.}
 analogously with the endpoints of the intervals given by
the corresponding quantiles of $F_0$ rather than those of $F_n$, i.e.
we use $[F_0^{-1}(\frac{j}{n}),F_0^{-1}(\frac{k}{n})]$ in place of
$[X_{(j)},X_{(k)}]$.
A simple counting argument shows that $\# \Japp(\ell) \leq 36\,\ell\,
2^{\ell}$, hence both $\Japp$ and $\Japp^0$ have a cardinality that
is bounded by $\sum_{\ell=1}^{\ell_{max}} 36\,\ell\,2^{\ell}$ $=O(n)$.
Thus both $P_n$ and $P_n^0$ can be computed in $O(n \log n)$ steps, where
the complexity is dominated by sorting the data. This advantage of
efficient computation plays an important in many applications.

By definition $\Japp(\ell)$ contains intervals whose empirical measure is
roughly the same, up to a factor of two. The `largest' intervals
at $\ell=2$ have empirical measure up to $\frac{1}{2}$; there is
no practical interest in considering larger intervals, and this upper
bound can be changed as appropriate. The `smallest' intervals at
$\ell=\ell_{max}$ have empirical measure of about $\log n /n$ since
in a density setting it is not possible to obtain consistent inference
with fewer observations. This particular choice of $\ell=\ell_{max}$
was also found to work well for the finite sample sizes
used in the simulation study in Section~\ref{simulations}.
The key parameter of the approximating set
is $d_{\ell}$, which describes how finely the endpoints are spaced
as a function of the length of the interval: Small intervals require
a fine spacing for a good approximation, while for large intervals
a coarser spacing is sufficient. The particular formula given by
$d_{\ell}$ ensures that intervals of all sizes are approximated
sufficiently well to guarantee optimal detection, as shown in
Theorem~\ref{detection},  while at the same
time the approximating set is sparse enough that one can control
$P_n$ simply with the union bound (this property does not hold
e.g. for the approximating set given in Rufibach and Walther~(2010)):

\begin{Proposition}   \label{boole}
Both $P_n$ and $P_n^0$ are $O_p(1)$ under $H_0$.
\end{Proposition}

Before proving Proposition~\ref{boole}, we note that the second
key ingredient besides the sparse approximating set
is the `standardization' of $F_n(I)$ in terms of the transformation
$\sqrt{2\lLR (F_0(I),F_n(I))}$ instead of the usual way to standardize
a binomial random variable. The latter case results in one tail that is not 
subgaussian and which is heavier than the other tail, see Shorack and 
Wellner~(1986,Ch.11.1), a problematic fact for the multiple testing
set-up considered here. In contrast, the `standardization'
via the above likelihood ratio transformation leads to clean
subgaussian tails: For a fixed interval $I$ and $t>0$
\be  \label{subgaussian}
\Pr\Bl(\sqrt{2\lLR (F_0(I),F_n(I))}>t\Br) \leq \exp\Bl(-\frac{1}{2}t^2\Br).
\ee 
While we could not find a statement of this result in the literature,
it is implicit in the proof of the Chernoff-Hoeffding theorem,
see Hoeffding~(1963): That proof establishes $\Pr(F_n(I) \geq v)
\leq \exp(-\lLR(F_0(I),v))$ for $v \in (F_0(I),1]$, see A.6.1
in van der Vaart and Wellner~(1996). Since it is easily seen that
the function $v \ra \lLR(F_0(I),v)$ is strictly increasing for $v>F_0(I)$, we
obtain $\Pr(\lLR(F_0(I),F_n(I)) >t) \leq \exp(-t)$, and (\ref{subgaussian})
follows. We note that (\ref{subgaussian}) also holds for the two-sided
version of the likelihood ratio provided one adds the factor 2 on the 
right side.

Since $\# \Japp^0(\ell) \leq 36\,\ell\, 2^{\ell}$ we obtain for $\kappa>2$:
\begin{equation*} 
\begin{split}
\Pr\Bl( \max_{I \in \Japp^0} &\Bl(\sqrt{2\lLR (F_0(I),F_n(I))}-
\sqrt{2 \log \frac{e}{F_0(I) \Bl(1-\min(F_n(I),\frac{1}{2})\Br)}}\Br) 
> \kappa \Br)  \\
& \leq \;\sum_{\ell =2}^{\ell_{max}} \# \Japp^0(\ell)\;
\max_{I \in \Japp^0(\ell)} 
\exp \Bl( -\frac{1}{2} \Bl( \sqrt{2 \log \frac{e}{F_0(I)}} +\kappa\Br)^2\Br)\\
& \leq \;\sum_{\ell =2}^{\ell_{max}} 36 \ell \;\exp \Bl( -\kappa \sqrt{\ell}-
\kappa^2/2\Br)\;\;\; \text{ since } F_0(I) \leq 2 \times 2^{-\ell} \\
& < \;C \exp (-\kappa^2/2) \\
\end{split}
\end{equation*}
for some universal $C>0$, proving Proposition~\ref{boole} for $P_n^0$,
but with $F_0(I)$ instead of $F_n(I)$ in the penalty term. Using
this result and (\ref{Tay}) one readily finds uniform bounds
on the ratios $F_n(I)/F_0(I)$ which allow to replace $F_0$ by
$F_n$ in the penalty term.

The proof of $P_n=O_p(1)$ is analogous, the main difference being
that the intervals $I$ are now random. Since by construction all intervals
$I \in \Japp$ have empirical measure at least $\log n/n$,
Lemma~\ref{tail} in Section~\ref{proofs} shows that
the tails of $\sqrt{2\lLR}$ are close enough to subgaussian
that the above argument goes through, concluding the proof of 
Proposition~\ref{boole}.

Finally we will also consider the direct penalization  of the scan
(\ref{Mn}), i.e. without approximating the set of all intervals:
$$
P_n^{all}\,:=\, \max_{\substack{
1\leq j<k\leq n\\
\log n \leq k-j \leq n/2}}
\Bl(\sqrt{2 \lLR\Bl(F_0([X_{(j)},X_{(k)}]),\frac{k-j+1}{n}\Br)}
-\sqrt{2 \log \frac{en^2}{(k-j)(n-k+j)}}\Br)
$$
Our main reason for investigating $P_n^{all}$ is that we need the 
following result for our theoretical analysis of
the average likelihood ratio in Section~\ref{ALR}:

\begin{Theorem}   \label{Pnall}
$P_n^{all}=O_p(1)$ under $H_0$.
\end{Theorem}

The restriction $k-j\geq \log n$ is necessary for this
result to hold since for very small intervals the tail of the test
statistic is far from subgaussian, causing the null distribution to blow
up, see Lemma~\ref{tail}. Of course, those small intervals are not
required for optimal detection, and $\Japp$ does not employ them either.

The proof of Theorem~\ref{Pnall} is neither short nor straightforward,
using the Hungarian construction. In contrast, the short proof of 
Proposition~\ref{boole}, given above, is essentially an application of Boole's
inequality together with a simple counting argument. This is one
of the two main advantages of using the approximating set $\Japp$,
the other being the
computational complexity of $O(n \log n)$, whereas $P_n^{all}$
requires to loop over $O(n^2)$ intervals.

Note that all versions of the scan introduced in this section are
distribution free and thus allow exact 
finite sample inference. The availability of algorithms with complexity
close to $O(n)$ is crucial for performing such a finite sample
inference, see Section~\ref{simulations} for details. 

The procedures in this section require the specification of $F_0$.
If $F_0$ is unknown, then these procedures can be viewed as goodness
of fit tests for some hypothesized $F_0$, with optimal power properties
against alternatives that concentrate more mass on some interval of
unknown location and length. It may also be possible to use these
procedures to construct confidence intervals for a distribution
function which improve on e.g. Kolmogorov-Smirnov bands.

\section{The condensed average likelihood ratio}  \label{ALR}

Chan and Walther~(2013) introduce the {\sl condensed average
likelihood ratio} in a regression setting and show that it allows
optimal detetion of a bump in a regression function. Here we investigate
its performance in a density context. Define
$$
A_{n}^{cond}\;:=\;\frac{1}{\# \Iapp} \sum_{I \in \Iapp} LR_n(F_0(I),F_n(I)),
$$
which is the average of the likelihood ratios $LR_n=\exp(\lLR)$ over
the approximating set of intervals

\begin{align*}
\Iapp &= \bigcup_{\ell =2}^{\ell_{max}} \Iapp(\ell),\;\;\; \mbox{ where }
\;\ell_{max}=\Bl\lfloor \log_2 \frac{n}{\log n}\Br\rfloor \;\;
\mbox{ and}\\
\Iapp(\ell) &= \Bl\{(X_{(j)},X_{(k)}]:
j,k \in \{1+i d_{\ell},
i=0,1,\ldots \}\ \mbox{ and } m_{\ell}<k-j\leq 2m_{\ell}\Br\},\\
&\;\;\;\;\mbox{ where } m_{\ell}=n2^{-\ell},\; d_{\ell}=
\Bl\lceil \frac{\sqrt{m_{\ell}}\,{\ell}^{4/5}}{\log n}\Br\rceil.
\end{align*}

Note that $\Iapp$ differs from $\Japp$ used above for the scan in
the choice of $d_{\ell}$. The different choice of $d_{\ell}$ is
necessary to guarantee optimal detection, but it still allows
computation in almost linear time since it is readily checked
that $\# \Iapp =O(n \log^2 n)$. A second difference is that
$\Iapp$ uses half-open intervals $(X_{(j)},X_{(k)}]$ rather
than closed intervals with a corresponding empirical measure
$\frac{k-j}{n}$ instead of $\frac{k-j+1}{n}$. These changes
guarantee that $A_{n}^{cond}$ will stay bounded under $H_0$:

\begin{Proposition}  \label{An}
$A_n^{cond}=O_p(1)$ under $H_0$.
\end{Proposition}

The density setting investigated here requires a proof that is more involved
than the one in the regression setting considered in
Chan and Walther~(2013). Further, in the density setting
there is no need to consider small intervals with empirical measure
less than about $\log n/n$, and $\Iapp$ is defined accordingly.

$A_n^{cond}$ is also distribution free and thus allows exact finite
sample inference.

\section{Optimality}  \label{optimality}

Next we investigate whether the penalized scans $P_n$ and $P_n^0$
and the condensed average likelihood ratio $A_n^{cond}$
allow optimal detection, i.e. whether they are able to detect
alternatives (\ref{density}) that satisfy
\be  \label{detectioncondition}
\sqrt{n} \frac{ F_{r,I}(I) -F_0(I)}{\sqrt{F_{r,I}(I)}}\; \geq \;
\sqrt{2 \log \frac{e}{F_{r,I}(I)}}\; (1+\eps_n),
\ee
with $\eps_n \sqrt{2 \log \frac{e}{F_{r,I}(I)}}
\ra \infty$. Note that both $r$ and $I$ may depend on $n$, but for
simplicity we will not include this in our notation.
Using arguments as in D\"{u}mbgen and Spokoiny~(2001)
and in Walther~(2010), one can show that no procedure can reliably
detect alternatives $F_{r,I}$ that satisfy (\ref{detectioncondition})
when $(1+\eps_n)$ is replaced by $(1-\eps_n)$. Thus
(\ref{detectioncondition}) does indeed describe a condition for
optimal detection. We note that while in the regression context
the `scale' of the effect is given by the spatial extent $|I|$, 
in the density context this role is played by the probability
$F_{r,I}(I)$.

\begin{Theorem}   \label{detection}
The penalized scans $P_n$ and $P_n^{all}$ and the condensed average 
likelihood ratio $A_n^{cond}$ provide optimal detection, i.e. they have 
asymptotic
power one uniformly in signals satisfying (\ref{detectioncondition}).
This result also holds for $P_n^0$ provided $F_0(I) > 2^{-\ell_{max}}$.
\end{Theorem}

Thus the optimality of $P_n^0$ comes with a proviso due to the
fact that the approximating set $\Japp^0$ is built from the null model
and not from the observed data: If the 
interval $I$ supporting the bump is very small, then the approximating
set $\Japp^0$ is not fine enough to allow optimal detection. While this
can be remedied by increasing $\ell_{max}$, such a step will severely affect
the computational complexity, and it is not clear a priori what an
appropriate choice for $\ell_{max}$ would be.
$P_n$ and $A_n^{cond}$ avoid this problem by using data-dependent
approximating sets. One of the consequences of Theorem~\ref{detection}
is that these approximating sets are rich enough for optimal detection
and there is no need to look over all intervals as in $P_n^{all}$.
This has obvious computational advantages as discussed above, and it
allows for a much simplified theoretical analysis: compare the proofs
of Proposition~\ref{boole} and Theorem~\ref{Pnall}. An interesting
distinction between the scan and the average likelihood ratio is
the fact that the approximating set will automatically lead
to optimal detection for the latter, but not for the former:
Evaluating the unpenalized scan $M_n$ on $\Japp$ or on the approximating sets 
given in Neill and Moore~(2004) or Arias-Castro et al.~(2005) will result in
optimal detection only on the smallest scales, i.e. for $F_{r,I}(I)
\approx 2 \frac{\log n}{n}$. Optimal detection on all
scales seems to require the use of scale-dependent critical values,
such as via a penalty term as in $P_n$ or via the {\sl blocked scan} introduced 
in Rufibach and Walther~(2010) and Walther~(2010).

\section{A simulation study}  \label{simulations}

We illustrate the theoretical results given above with a simulation
study that compares the performance of the scan, the penalized scan $P_n$,
and the condensed average likelihood ratio $A_n^{cond}$. In order to
arrive at a fair comparison, we evaluate the scan $M_n$ only over intervals
that contain between $\log n$ and $n/2$ observations. This increases the 
power of the scan compared to the original definition (\ref{scan}) and provides
the same a priori assumptions about the length of the cluster for all
three methods.

Note that since $F_0$ is known we may assume that $F_0$ is the
$U[0,1]$ distribution: Applying the transformation $Y=F_0(X)$
transforms the model (\ref{density}) into
\be \label{density2}
f_{r,I}(y)\;=\; \frac{r 1(y \in I) + 1(y \in I^c)}{r|I|+1-|I|}\;1(y \in [0,1]),
\ee
where the interval $I$ is the image of the original interval $I$
under the map $F_0$. Moreover, all the statistics $M_n, P_n, P_n^0, P_n^{all}$,
and $A_n^{cond}$ are seen to be distribution free. Hence we may
simulate the null distributions
of these statistics by drawing $X_1,\ldots,X_n$ i.i.d. U[0,1] (say), thus
allowing for exact (up to Monte Carlo simulation error) finite sample
inference.

Tables~\ref{table1} and~\ref{table2} list the power at the 5\% significance
level for sample
sizes $n=10^4$ and $n=10^6$, respectively. Each case considers the
range for the effect ratio $r$ where detection starts to become possible,
for a small interval and for a large interval $I$. These simulations 
illustrate how the optimality result of Section~\ref{optimality} about  
$P_n$ and $A_n^{cond}$ sets in. In contrast, one sees that the scan
$M_n$ is competitive only for signals on the smallest scales and it is
inferior to $P_n$ and $A_n^{cond}$ otherwise. In the context of regression,
the inferiority of the scan at larger scales was expounded theoretically
by Chan and Walther~(2013).
Note that unlike in the regression context, `scale' is not given by the
length $|I|$ but by $F_{r,I}(I)$, which is of the order $rF_0(I)$ as long as 
the latter quantity stays bounded.

The simulations show that the condensed average likelihood ratio
$A_n^{cond}$ has arguably the best overall performance among the three
procedures considered.

\begin{table}[h]
\begin{tabular}{c|ccc||c|ccc}
\multicolumn{4}{c||}{$|I|=10^{-3}$} & \multicolumn{4}{c}{$|I|=0.3$} \\ \hline 
r & scan & pen.scan & cond.ALR & r & scan & pen.scan & cond.ALR \\ \hline 
1.8 &  09 & 07 &   05 & 1.01 & 05 &  06 & 08 \\
2.1 &  15 & 14 &   11 & 1.03 & 06 &  10 & 18 \\
2.4 &  31 & 24 &   22 & 1.05 & 09 &  23 & 39 \\
2.7 &  46 & 48 &   36 & 1.07 & 17 &  47 & 70 \\
3.0 &  67 & 65 &   60 & 1.09 & 37 &  79 & 90 \\
3.3 &  82 & 79 &   74 & 1.11 & 66 &  92 & 97 \\
3.6 &  92 & 92 &   85 & 1.13 & 89 &  99 & 100 \\
3.9 &  97 & 97 &   94 & 1.15 & 97 & 100 & 100 \\
4.2 &  99 & 99 &   98 &      &    &     &
\end{tabular}
\caption{Power in percent for detecting clusters (\ref{density})
for various values of $r$ and two different lengths of $I$
for sample size $n=10^4$. }
\label{table1}
\end{table}

\begin{table}[h]
\begin{tabular}{c|ccc||c|ccc}
\multicolumn{4}{c||}{$|I|=10^{-4}$} & \multicolumn{4}{c}{$|I|=0.3$} \\ \hline
r & scan & pen.scan & cond.ALR & r & scan & pen.scan & cond.ALR \\ \hline
1.25 & 06 & 06 & 05 & 1.002 & 06 & 07 & 10 \\
1.35 & 07 & 08 & 07 & 1.004 & 05 & 14 & 23 \\
1.45 & 14 & 16 & 15 & 1.006 & 05 & 38 & 52 \\
1.55 & 35 & 40 & 34 & 1.008 & 09 & 69 & 80 \\
1.65 & 61 & 66 & 62 & 1.010 & 14 & 91 & 96 \\
1.75 & 83 & 86 & 85 & 1.012 & 39 & 99 & 99 \\
1.85 & 96 & 97 & 95 & 1.014 & 71 & 100 & 100 \\
1.95 & 99 & 99 & 99 & 1.016 & 92 & 100 & 100 \\
     &    &    &    & 1.018 & 99 & 100 & 100 \\
\end{tabular}
\caption{Power in percent for detecting clusters (\ref{density})
for various values of $r$ and two different lengths of $I$
for sample size $n=10^6$. }
\label{table2}
\end{table}

In the above simulations the finite sample exact critical values and the
power were
approximated with $10^5$ and $10^3$ simulations, respectively.
The location of the interval $I$ was randomized in each simulation
to avoid confounding the results with the particular construction of
the approximating sets $\Iapp$ and $\Japp$.
In the case of the sample size $n=10^6$, the scan $M_n$ was evaluated on
the approximation set $\Japp$, i.e. the same approximation set used for
$P_n$, since looking at all intervals was computationally prohibitive.
Conversely, for sample size $n=10^4$ we examined the effect of the 
approximating set by running the simulation with the
penalized scan and the condensed average likelihood ratio evaluated
over all intervals containing between $\log n$ and $n/2$ observations
rather than evaluating them over an approximating set. We observed only
very small changes in power, mostly decreases, and the computation time was 
much longer. This confirms the theoretical finding from Section~\ref{optimality}
that it suffices the evaluate these statistics over an approximating set,
which yields tremendous computational advantages without sacrificing
detection power.

\section{Proofs} \label{proofs}

\subsection{Preliminary results} 

\nin {\bf 1.} Using $\log x \leq x-1$ and a two term Taylor expansion, 
respectively, gives for $a,b \in (0,1)$:
\be  \label{Tay}
\begin{split}
n \frac{(b-a)^2}{a(1-a)}\; \geq \; \lLR(a,b) \;&= \;\frac{n}{2 \xi (1-\xi)}
(b-a)^2 1(a<b)\;\;\;\text{ for $\xi \in (a,b)$}\\
& \; \geq \frac{n}{2 b}(b-a)^2 1(a<b) \\
\end{split}
\ee

\nin {\bf 2.} Let $I$ be an interval satisfying 
$\ell:= \lfloor
\log_2 1/F_n(I) \rfloor +1 \leq \ell_{max}$, so 
$m_{\ell} <nF_n(I) \leq 2m_{\ell}$.
Then by construction of $\Japp(\ell)$ there exists 
$\tilde{I} \in \Japp (\ell)$
such that 
\be  \label{AP}
F_n (I \triangle \tilde{I}) \; \leq \;2\frac{d_{\ell}-1}{n}
\; \leq \; \frac{F_n(I)}{3\sqrt{\ell}},
\ee
and the same result holds for $\Japp^0$ with $F_n$ replaced by $F_0$ in the 
above.

\begin{Lemma}  \label{approx}
Let $J$ be an interval and $F_{r,I}$ be the distribution given in 
(\ref{density}) with $r\geq 1$. Then for $G \in \{F_0,F_{r,I}\}$:
\begin{gather*}
(F_{r,I}-F_0)(J)  \;\geq \;
(F_{r,I}-F_0)(I) \Bl(1-\frac{G(I \triangle J)}{G(I)}\Bl)\;\;\;
\text{ if } G(I) \leq \frac{1}{2},\;\text{ and}\\
1-\frac{F_0(I \setminus J)}{F_0(I)} \;\leq \; \frac{F_{r,I}(J)}{F_{r,I}(I)}
\;\leq \;1+\frac{F_0(J \setminus I)}{F_0(I)}.
\end{gather*}
\end{Lemma}

For a proof of Lemma~\ref{approx} note that
$$
f_{r,I}(x)-f_0(x) =
\begin{cases}
d_{r,I} f_0(x)\;\;\text{ if } x \in I\\
-\frac{F_0(I)}{F_0(I^c)} d_{r,I} f_0(x)\;\;\text{ if } x \in I^c\\
\end{cases}
$$
where $d_{r,I}:=r/(rF_0(I)+F_0(I^c))-1\geq 0$ since $r\geq 1$.
Hence
\begin{align*}
(F_{r,I}-F_0)(J) &= d_{r,I} F_0(I \cap J) -\frac{F_0(I)}{F_0(I^c)} d_{r,I}
F_0(J\setminus I)\\
& =(F_{r,I}-F_0)(I) \Bigl( \frac{F_0(I \cap J)}{F_0(I)} -
\frac{F_0(J \setminus I)}{F_0(I^c)}\Br)\\
& \geq (F_{r,I}-F_0)(I) \frac{F_0(I \cap J) -F_0(J \setminus I)}{F_0(I)}
\;\;\;\text{ if } F_0(I) \leq \frac{1}{2}
\end{align*}
and the claim for $G=F_0$ follows by writing $F_0(I \cap J)=F_0(I)-
F_0(I \setminus J)$. The claim for $G=F_{r,I}$ follows since
$\frac{F_0(I \cap J)}{F_0(I)}-\frac{F_0(J \setminus I)}{F_0(I^c)}=
\frac{F_{r,I}(I \cap J)}{F_{r,I}(I)}-\frac{F_{r,I}(J \setminus 
I)}{F_{r,I}(I^c)}$
by the definition of $f_{r,I}$. The second claim follows from
dividing the inequality $F_{r,I}(I) -F_{r,I}(I \setminus J)
\leq F_{r,I}(J) \leq F_{r,I}(I)+F_{r,I}((J\setminus I)$ by
$F_{r,I}(I)$ and observing $\frac{F_{r,I}(I \setminus J)}{
F_{r,I}(I)} = \frac{F_0(I \setminus J)}{F_0(I)}$ and
$\frac{F_{r,I}(J \setminus I)}{F_{r,I}(I)} =\frac{1}{r}
\frac{F_0(J \setminus I)}{F_0(I)}$ by the definition of $f_{r,I}$.
$\hfill \Box$

The following lemma is an extension of Proposition~2.1 in
D\"{u}mbgen~(1998):
\begin{Lemma} \label{tail}
Denote the two-sided log likelihood ratio statistic by 
$\lLR^{two}(a,b):=nb\log \frac{b}{a}+n(1-b)\log \frac{1-b}{1-a}$
and the one sided versions by $\lLR^{left}(a,b):=\lLR^{two}(a,b) 1(a<b)$
and by $\lLR^{right}(a,b):=\lLR^{two}(a,b) 1(a>b)$.
Hence $\lLR^{left}$ equals $\lLR$ used above. 
Let $U_1,\ldots,U_n$ be i.i.d. $U[0,1]$, so $U_{(k)}-U_{(j)} \sim
$beta$(k-j,n+1-k+j)$ for $1 \leq j < k \leq n$. Set 
$p_{jk}:=\frac{k-j}{n+1}$. Then for $p \in (0,1)$ and $t>0$:
\begin{align*}
\Pr \Bl( \sqrt{2 \lLR^{two}(U_{(k)}-U_{(j)},p)}>t\Br) & \leq
2 \exp \Bl\{-\min \Bl(\frac{p_{jk}}{p}, \frac{1-p_{jk}}{1-p}\Br)
\frac{(n+1)}{n} \frac{t^2}{2} +n\frac{p-p_{jk}}{1(p>p_{jk})-p_{jk}}\Br\} \\
& \leq 
\begin{cases}
2 \exp \Bl(-\frac{t^2}{2}\Br)& \text{ if }p:=p_{jk} \\
2 \exp \Bl(-\frac{(k-j)}{(k-j+1)} \frac{t^2}{2} +3\Br)
& \text{ if }p:=\frac{k-j+1}{n} \leq \frac{1}{2}.
\end{cases} 
\end{align*}
In more detail:
\begin{align*}
\Pr \Bl( \lLR^{left}(U_{(k)}-U_{(j)},p)>t\Br) 
& \leq \exp \Bl\{-\frac{p_{jk}}{p} \frac{(n+1)}{n} \Bl(t-
n\frac{(p-p_{jk})(p-p_{jk}1(p_{jk}>p))}{p_{jk}(1-p_{jk})}\Br)\Br\}\\
\Pr \Bl( \lLR^{right}(U_{(k)}-U_{(j)},p)>t\Br)
& \leq \exp \Bl\{-\frac{(1-p_{jk})}{(1-p)} \frac{(n+1)}{n} \Bl(t-
n\frac{(p_{jk}-p)[1-p-(1-p_{jk})1(p_{jk}<p)]}{p_{jk}(1-p_{jk})}\Br)\Br\}
\end{align*}
\end{Lemma}

Hence in the case of random intervals whose lengths follow the beta
distribution, $\sqrt{2 \lLR^{two}}$ has subgaussian tails for $p=p_{jk}$.
For $p$ close to $p_{jk}$ the tails are still subgaussian but with
a scale factor that is smaller in one tail and larger in the other.

{\bf Proof of Lemma~\ref{tail}:} For $u \in (0,p)$:
\be  \nn
\begin{split}
\lLR^{left} & (u,p) = \lLR^{two}(u,p) \\ 
& = \frac{p}{p_{jk}} \lLR^{two}(u,p_{jk}) +\lLR^{two}(p_{jk},p)
  +n\frac{p_{jk}-p}{p_{jk}} \log \frac{1-p_{jk}}{1-u} \\
& \leq \frac{p}{p_{jk}} \lLR^{two}(u,p_{jk}) +n\frac{(p_{jk}-p)^2}{p_{jk}
(1-p_{jk})} +n\frac{p_{jk}-p}{p_{jk}} \log(1-p_{jk})\;1(p_{jk}<p)\;\;\;
\text{ by (\ref{Tay})}\\
& \leq \frac{p}{p_{jk}} \lLR^{left}(u,p_{jk}) +n\frac{(p-p_{jk})(p-
p_{jk}1(p_{jk}>p))}{p_{jk}(1-p_{jk})}
\end{split}
\ee
since $-(1-p_{jk}) \log(1-p_{jk}) \leq p_{jk}$, and because
$\lLR^{left}(u,p)$ is non-increasing in $u$ while
$\lLR^{two}(u,p_{jk})$ is increasing for $u>p_{jk}$. Hence the inequality
above must also hold with $\lLR^{two}(u,p_{jk})$ replaced by $\min \Bl(
\lLR^{two}(u,p_{jk}), \lLR^{two}(p_{jk},p_{jk}) \Br)=\lLR^{left}(u,p_{jk})$.
 Now the probability inequality for $\lLR^{left}$
follows from the above inequality together with $\Pr \Bl( \lLR^{left}(U_{(k)}
-U_{(j)}, p_{jk}) >t\Br) \leq \exp \Bl(-\frac{n+1}{n}t \Br)$,
which is a consequence of Proposition~2.1 in D\"{u}mbgen~(1998).
The inequality for the right tail follows analogously, and the
tail bound for $\sqrt{2 \lLR^{two}}$ obtains as a consequence. $\hfill \Box$

\subsection{Proof of Theorem~\ref{Pnall}}  \label{Pnallproof}

Under $H_0$, $(F_0(X_1),\ldots,F_0(X_n)) \stackrel{d}{=}
(U_1,\ldots,U_n)$, where $U_1,\ldots,U_n$ are i.i.d. U[0,1].
We divide the collection of intervals under consideration into
a collection of small intervals 
$$
{\cal S}_n:=\Bl\{(j,k):\;1 \leq j<k\leq n,\; \log n \leq k-j \leq \log^2n \Br\}
$$
and the collection of the remaining intervals
$$
{\cal I}_n:=\Bl\{(j,k);\;1 \leq j<k\leq n,\; \log^2n <k-j \leq n/2\Br\}.
$$ 
The cardinality of ${\cal S}_n$ is small enough that we can use the
union bound to show
\be \label{1}
\max_{(j,k) \in {\cal S}_n} \Bigg(\sqrt{2 \lLR\Bl(U_{(k)}-U_{(j)},
\frac{k-j+1}{n}\Br)} -\sqrt{2 \log \frac{en^2}{(k-j)(n-k+j)}}\Bigg)^+
=o_p(1)
\ee
For the larger intervals we approximate $\sqrt{2\lLR}$
by the normalized increment of the uniform quantile process:
\be  \label{2}
\max_{(j,k) \in {\cal I}_n} \Bigg|\sqrt{2 \lLR\Bl(U_{(k)}-U{(j)},
\frac{k-j+1}{n}\Br)} -\sqrt{n}\frac{\Bl|\frac{k-j}{n}-\Bl(U_{(k)}-U{(j)}\Br)
\Br|}{\sqrt{\frac{k-j}{n}\Bl(1-\frac{k-j}{n}\Br)}} \Bigg|=O_p(1)
\ee

The normalized increments of the uniform  quantile process can
in turn be approximated on ${\cal I}_n$ by the normalized
increments of a Brownian bridge $B$:

\be \label{3}
\max_{(j,k) \in {\cal I}_n} \Bigg| \sqrt{n}\frac{\Bl|\frac{k-j}{n}
-\Bl(U_{(k)}-U{(j)}\Br)\Br|}{\sqrt{\frac{k-j}{n}\Bl(1-\frac{k-j}{n}\Br)}}
-\frac{\Bl|B(\frac{k}{n})-B(\frac{j}{n})\Br|}{\sqrt{\frac{k-j}{n}\Bl(1-
\frac{k-j}{n}\Br)}} \Bigg|=O_p(1)
\ee

Theorem~\ref{Pnall} follows from (\ref{1}--\ref{3}) together with
\be  \label{4}
\sup_{0<s<t<1} \Bl(\frac{|B(t)-B(s)|}{\sqrt{(t-s)(1-(t-s))}}-
\sqrt{2 \log \frac{e}{(t-s)(1-(t-s))}} \Br) < \infty\;a.s.
\ee

The above results also show how one might design
an appropriate penalty if one wishes to scan over very large
intervals, i.e. $(k-j)/n$ close to 1. Indeed, it is well known that
the normalized uniform quantile process blows up both at 0 and
at 1, see Ch. 16 in Shorack and Wellner~(1986).

{\bf Proof of (\ref{1})}: Clearly $\# {\cal S}_n \leq n \log^2 n$. Hence
the tail inequality for $\sqrt{2 \lLR}$ given in Lemma~\ref{tail} yields
for $C>0$:
\be \nn
\begin{split}
\Pr & \Bigg( \max_{(j,k) \in {\cal S}_n} \Bigg( \sqrt{2 \lLR \Bl(U_{(k)}
-U_{(j)},
\frac{k-j+1}{n}\Br) } -\sqrt{2 \log \frac{en^2}{(k-j)(n-k+j)}}\Bigg)
>C \Bigg) \\
& \leq n (\log^2 n) \max_{(j,k) \in {\cal S}_n} 2 e^3 \exp \Bigg\{
-\frac{k-j}{2(k-j+1)} \Bigg(C+\sqrt{2 \log \frac{en^2}{(k-j)(n-k+j)}}\Bigg)^2
\Bigg\} \\
& \leq 2 e^3 n (\log^2 n) \exp \Bl\{ -\Bl(1-\frac{1}{\log n}\Br)
\Bl( \frac{C^2}{2} +\log \frac{en}{\log^2n} +C\sqrt{2 \log \frac{en}{
\log^2n}}\Br) \Br\}\\
& \,\,\,\,\,\,\text{ since $(k-j)(n-k+j) \leq n \log^2 n\;$ on ${\cal S}_n$}\\
& \leq 2 e^3 (\log^4 n) \exp \Bl\{ -\Bl(1-\frac{1}{\log n}\Br)
\Bl( \frac{C^2}{2} +C\sqrt{2 \log \frac{en}{
\log^2n}}\Br) \Br\} \; \ra 0
\end{split}
\ee

{\bf Proof of (\ref{2})}: By (\ref{Tay})
\begin{align}  
\Bigg| &\sqrt{2 \lLR \Bl(U_{(k)} -U_{(j)},\frac{k-j+1}{n}\Br) } -\sqrt{n}
\frac{\Bl| \frac{k-j+1}{n} -(U_{(k)} -U_{(j)})\Br|}{\sqrt{
\frac{k-j}{n} \Bl(1-\frac{k-j}{n}\Br)}} \Bigg| \nn \\
& = \sqrt{n} \Bl| \frac{k-j+1}{n} -(U_{(k)} -U_{(j)})\Br|
\Bigg| \sqrt{\frac{1}{\xi (1-\xi)}} - \sqrt{\frac{1}{
\frac{k-j}{n} \Bl(1-\frac{k-j}{n}\Br)}} \Bigg| \label{2st}
\end{align}

for $\xi$ between $\frac{k-j+1}{n}$ and $U_{(k)} -U_{(j)}$.
On the event 
$$
{\cal A}_n(C):=\Bl\{\Bl|U_{(k)} -U_{(j)} -\frac{k-j}{n}\Br|
\leq \Bl(C+\sqrt{2 \log \frac{en^2}{(k-j)(n-k+j)}}\Br)
\sqrt{\frac{k-j}{n^2} \Bl(1-\frac{k-j}{n}\Br)}\,\,
\text{ for all }(j,k) \in {\cal I}_n\Br\}
$$
 we have $\Bl|
U_{(k)} -U_{(j)} -\frac{k-j}{n}\Br| \leq \frac{ (C +\sqrt{2 \log n})}{
\sqrt{k-j}} \frac{(k-j)}{n} \leq \frac{2}{\sqrt{ \log n}} \frac{(k-j)}{
n}$ eventually. Hence
\begin{align*}
\Bigg| \frac{1}{\xi (1-\xi)} -\frac{1}{\frac{k-j}{n} \Bl(1-\frac{k-j}{n}\Br)}
\Bigg| & \leq \frac{1}{\xi \frac{k-j}{n}} \bigg|\xi -\frac{k-j}{n} \bigg|
+\frac{1}{(1-\xi)(1-\frac{k-j}{n})} \bigg|\xi -\frac{k-j}{n} \bigg| \\
& \leq \frac{4n}{(k-j)\sqrt{\log n}} + \frac{4n}{(n-k+j)\sqrt{\log n}}\\
& = \frac{4}{\frac{k-j}{n} \Bl(1-\frac{k-j}{n}\Br) \sqrt{\log n}}.
\end{align*}

Since $0 < b-a < b/2$ for $a,b>0$ implies $|\sqrt{b}-\sqrt{a}|
\leq (b-a)/\sqrt{b}$, (\ref{2st}) is not larger than
\begin{multline}  \nn
\frac{4 \sqrt{n} \Bl|\frac{k-j+1}{n} -(U_{(k)} -U_{(j)})\Br|}{
\sqrt{\frac{k-j}{n} \Bl(1-\frac{k-j}{n}\Br)\log n}}
\; \leq \; \frac{4\Bl(C+\sqrt{2 \log \frac{en^2}{(k-j)(n-k+j)}}\Br)}{
\sqrt{ \log n}}  \\
 + \frac{4}{\sqrt{n}
\sqrt{\frac{k-j}{n} \Bl(1-\frac{k-j}{n}\Br)\log n}}
\; \leq \;4 \frac{C+\sqrt{2 \log n}}{\sqrt{ \log n}} + \frac{8}{
(\log n)^{3/2}} \;\;\;\text{ for } (j,k) \in {\cal I}_n.
\end{multline}

(\ref{2}) follows since $\lim_{C \ra \infty} \liminf_{n \ra \infty}
\Pr ({\cal A}_n(C))=1$ by (\ref{3}) and (\ref{4}), and replacing
$\frac{k-j}{n}$ with $\frac{k-j+1}{n}$ in the numerator of the second
term in (\ref{2}) incurs a difference bounded by $8/\log n$ as seen
above.

{\bf Proof of (\ref{3})}:
By the Hungarian construction, see Theorem 12.2.2 in Shorack and
Wellner~(1986), there exists a version $B_n$ of the Brownian
bridge on $[0,1]$ such that
\begin{align*}
\limsup_n & \max_{(j,k) \in {\cal I}_n} \bigg| \sqrt{n} \Bl|
\frac{k-j}{n}-(U_{(k)} -U_{(j)})\Br|-\Bl|B_n(\frac{k}{n})-
B_n(\frac{k}{n})\Br|\bigg|\\
& \leq \limsup_n \max_{(j,k) \in {\cal I}_n}\bigg(
\Bl| \sqrt{n} \Bl(U_{(k)} -\frac{k}{n}\Br)-B_n(\frac{k}{n})\Br|
+\Bl|\sqrt{n} \Bl(U_{(j)} -\frac{j}{n}\Br)-B_n(\frac{j}{n})\Br| \bigg)\\
& \leq 2M \frac{\log n}{\sqrt{n}} \;\; \text{ a.s. form some } M< \infty
\end{align*}

The claim follows since $\sqrt{\frac{k-j}{n} \Bl(1-\frac{k-j}{n}\Br)}
\geq \frac{\log n}{2\sqrt{n}}$ for $(j,k) \in {\cal I}_n$.

{\bf Proof of (\ref{4})}:
This can be proved using Theorem~6.1 in D\"{u}mbgen and Spokoiny~(2001):
On the set of all intervals ${\cal T}:=\{(s,t]:0 <s<t<1\}$ define
the metric $\rho$ via $\rho^2\Bl((s,t],(s',t']\Br):=|s-s'|+|t-t'|$
and the stochastic process $X\Bl((s,t]\Br):=B(t)-B(s)$. With 
$\sigma^2 \Bl((s,t]\Br):=(t-s)(1-t+s)$ one readily checks that
$\sigma^2\Bl((s,t]\Br) \leq \sigma^2\Bl((s',t']\Br) +\rho^2 \Bl((s,t],
(s',t']\Br)$. Since $X\Bl((s,t]\Br)/\sigma \Bl((s,t]\Br) \sim $N(0,1),
the subgaussian tail condition (i) of said theorem holds. To prove the
subgaussian tail condition (ii) for the variation of $X$, write
$B(t)=W(t)-tW(1)$ for a Brownian motion $W$. Then
\begin{multline} \nn
\frac{X\Bl((s,t]\Br)-X\Bl((s',t']\Br)}{\rho\Bl((s,t],(s',t']\Br)}
=\frac{W\Bl((s,t]\setminus (s',t']\Br)-W\Bl((s',t']\setminus (s,t]\Br)}{
\sqrt{|s-s'|+|t-t'|}} -W(1) \frac{(t-s)-(t'-s')}{\sqrt{|s-s'|+|t-t'|}}\\
\stackrel{d}{=} N\Bigg( 0,\frac{\text{Leb} \Bl((s,t] \triangle (s',t']\Br)}{
|s-s'|+|t-t'|} + \frac{\Bl( (t-s) -(t'-s')\Br)^2}{|s-s'|+|t-t'|}-
2\frac{\text{Leb} \Bl((s,t] \triangle (s',t']\Br) \Bl((t-s)-(t'-s')\Br)}{
|s-s'|+|t-t'|} \Bigg)
\end{multline}

The latter variance is not more than four, hence condition (ii) holds
with $L=1$ and $M=8$. Finally, a calculation similar as in
D\"{u}mbgen and Spokoiny~(2001) shows that $V=1$. (\ref{4}) follows.

Checking condition (iii), i.e. establishing an exponential inequality
for the variation of the process under consideration, is the technically 
most challenging aspect in applying said Theorem~6.1, see also e.g.
D\"{u}mbgen and Walther~(2008).
Here we approached this problem via the Hungarian construction, which leads
to the simpler task of establishing an exponential inequality for the 
variation of the increments of a Brownian bridge. $\hfill \Box$

\subsection{Proof of Proposition~\ref{An}}

We use $F_0\Bl((X_{(j)},X_{(k)}]\Br) \stackrel{d}{=} U_{(k)}-U_{(j)}$
for $U_1,\ldots,U_n$ i.i.d. U[0,1] and define the event
$$
{\cal B}_{m,n}:=\Bl\{ \sqrt{2 \lLR \Bl(U_{(k)}-U_{(j)},\frac{k-j}{n}
\Br)}
\leq \sqrt{2 \log \frac{en^2}{(k-j)(n-k+j)}} +m  \text{ for all }
(j,k) \in \Iapp \Br\}.
$$
We will show that for $(j,k) \in \Iapp (\ell)$
\be  \label{15st}
\Ex \Bl( LR_n \Bl(U_{(k)}-U_{(j)},\frac{k-j}{n} \Br) 1_{{\cal B}_{m,n}}
\Br) \;\leq \; 14(\sqrt{2\ell} +m)\;\;\text{ eventually, uniformly in}
\ee
$(j,k)$ and $\ell$.
Then $A_n^{cond}=O_p(1)$ can be shown as in the proof of Theorem~3 in 
Chan and Walther~(2013) since $\lim_{m \ra \infty} \liminf_{n \ra \infty}
\Pr({\cal B}_{m,n})=1$ by Theorem~\ref{Pnall}, which is readily seen
to hold also with $\frac{k-j}{n}$ in place of $\frac{k-j+1}{n}$
in the definition of $P_n^{all}$.

To prove (\ref{15st}) fix $(j,k) \in \Iapp (\ell)$. We will show below
that on the event ${\cal B}_{m,n}$ for $n \geq n_0(m)$
\begin{enumerate}
\item[(a)] $u:=U_{(k)}-U_{(j)}$ falls in an interval $B$ of length
at most 
$$
4 \sqrt{\frac{\frac{k-j}{n}\Bl(1-\frac{k-j}{n} \Br)}{n}} \Bl(
C\Bl(\frac{k-j}{n}\Br) +m \Br),\;\;\text{ where } C(\delta):=
\sqrt{2 \log \frac{1}{\delta}},
$$
and
\item[(b)] $u \geq \frac{k-j}{8n}$.
\end{enumerate}

Using the fact that $U_{(k)}-U_{(j)} \sim$ beta$(k-j,n+1-k+j)$ we
can then compute
\begin{align*}
\Ex \Bl( LR_n & \Bl(U_{(k)}-U_{(j)},\frac{k-j}{n} \Br) 1_{{\cal B}_{m,n}}
\Br)\\
& = \int_B \Bl(\frac{k-j}{nu}\Br)^{k-j} \Bl(\frac{1-\frac{k-j}{n}}{
1-u}\Br)^{n-k+j} u^{k+j-1} (1-u)^{n-k+j}
\frac{n!}{(k-j-1)! (n-k+j)!} \;du \\
& \leq  \frac{e}{2 \pi} \int_B \frac{k-j}{nu} \sqrt{\frac{n}{
\frac{k-j}{n} \Bl(1-\frac{k-j}{n} \Br)}} \;du \;\;\;\text{ by Stirling's
formula}\\
& \leq \frac{e}{2 \pi} 32 \Bl(C\Bl(\frac{k-j}{n}\Br) +m \Br)
\end{align*}
by (a) and (b). (\ref{15st}) follows since $(j,k) \in \Iapp (\ell)$
implies $\frac{k-j}{n} > 2^{-\ell}$.

(a) follows for $n \geq n_0(m)$ from the inequality given in
Proposition~2.1 in D\"{u}mbgen~(1998) together with the fact that
$k-j \geq \log n$ by the construction of $\Iapp$. Said inequality
yields in particular
$$
u \;\geq \;\frac{k-j}{n} - \sqrt{\frac{\frac{k-j}{n}\Bl(1-
\frac{k-j}{n} \Br)}{n}}
\Bl(C\Bl(\frac{k-j}{n}\Br) +m \Br) \;\geq \;
\frac{k-j}{n} \Bl(1-\frac{C\Bl(\frac{k-j}{n}\Br) +m}{\sqrt{k-j}}\Br).
$$
Thus in the case $k-j \geq 4 \log n$, (b) follows since $\sqrt{k-j} \geq
\frac{8}{7} \Bl(C\Bl(\frac{k-j}{n}\Br) +m \Br)$ for $n \geq n_0(m)$.
In the case $k-j=b \log n$ with $b \in[1,4)$, consider $u:=\frac{k-j}{8n}$.
Then a standard calculation shows that $\lLR \Bl(u,\frac{k-j}{n}\Br)
\geq \Bl( \frac{9}{8} +o(1)\Br) \log n$, where the $o(1)$ term is
uniform in $b$. Thus this choice of $u$ violates the inequality
defining ${\cal B}_{m,n}$ for $n \geq n_0(m)$. Since 
$\lLR \Bl(u,\frac{k-j}{n}\Br)$ increases as $u$ moves away from 
$\frac{k-j}{n}$, this implies that we must have $u>\frac{k-j}{8n}$
for $n \geq n_0(m)$, completing the proof of (b). $\hfill \Box$

\subsection{Proof of Theorem~\ref{detection}}

We first prove the claim about $P_n^0$. Consider an alternative
(\ref{density}) that satisfies (\ref{detectioncondition}) and
also $F_0(I) > 2^{-\ell_{max}}$. Then $\ell:= \lfloor \log_2 1/F_0(I) 
\rfloor +1 \leq \ell_{max}$, so by (\ref{AP}) there exists
$\tilde{I} \in \Japp^0 (\ell)$ with $F_0 (I \triangle \tilde{I})
\leq \frac{F_0(I)}{3\sqrt{\ell}}$. 
Set $b_n:=\eps_n \sqrt{2 \log \frac{e}{F_{r,I}(I)}}$,
so $b_n \ra \infty$ by assumption (\ref{detectioncondition}).
On the event ${\cal A}_n:=
\Bl\{F_n(\tilde{I}) \geq F_{r,I}(\tilde{I}) -\sqrt{\frac{
F_{r,I}(\tilde{I}) b_n}{n}}\Br\}$ condition (\ref{detectioncondition})
implies $F_n(\tilde{I}) \geq F_0(\tilde{I})$ and hence
\begin{equation*}
\begin{split}
\sqrt{2\lLR (F_0(\tilde{I}),F_n(\tilde{I}))} \; &\geq \; 
\sqrt{n}\frac{F_n(\tilde{I})-
F_0(\tilde{I})}{\sqrt{F_n(\tilde{I})}} \;\;\text{ by (\ref{Tay})}\\
& \geq \;\sqrt{n}\frac{F_{r,I}(\tilde{I})-
F_0(\tilde{I})}{\sqrt{F_{r,I}(\tilde{I})}}-\sqrt{b_n}
 \;\;\text{ on ${\cal A}_n$ since }x \ra \frac{x-F_0(\tilde{I})}{x} \nearrow \\
& \geq \;\sqrt{n}\frac{F_{r,I}(I)-F_0(I)}{\sqrt{F_{r,I}(I)}}
\Bl(1-\frac{1}{3\sqrt{\ell}}\Br)^2 -\sqrt{b_n}
 \;\;\text{ by Lemma~\ref{approx}}\\
& \geq \;\Bl(\sqrt{2 \log \frac{e}{F_{r,I}(I)}}+b_n\Br)
\Bl(1-\frac{2}{3\sqrt{\log \frac{e}{F_{r,I}(I)}}}\Br) 
-\sqrt{b_n}\\
& \geq \;\sqrt{2 \log \frac{e}{3F_n(\tilde{I})}} +\frac{1}{3}b_n
 -1-\sqrt{b_n}\\
\end{split}
\end{equation*}
where the last inequality hold by Lemma~\ref{approx} and on the event 
${\cal B}_n:=\Bl\{F_{r,I}(\tilde{I})\leq 2F_n(\tilde{I})\Br\}$.
Chebyshev's inequality gives $\Pr({\cal A}_n) \geq 1-\frac{1}{b_n}$ and
$\Pr({\cal B}_n) \geq 1-\frac{4}{nF_{r,I}(\tilde{I})} \geq 
1-\frac{3}{\log n}$,
where the last inequality follows with Lemma~\ref{approx} from
$F_{r,I}(I) \geq 2 \log n/n$, which in turn is a consequence
of (\ref{detectioncondition}).
Hence $P_n^0 \ra \infty$ with probability converging to 1, uniformly
in alternatives satisfying (\ref{detectioncondition}). On the other
hand, the critical value of $P_n^0$ stays bounded by Proposition~\ref{boole}.

To prove the claim for $P_n$ note that by (\ref{AP}) we can find
$\tilde{I} \in \Japp (\ell)$ such that $F_n(I \triangle \tilde{I})
\leq \frac{F_n(I)}{3\sqrt{\ell}}$ by taking $\ell
:= \lfloor \log_2 1/F_n(I) \rfloor +1$. This index satisfies
$\ell \leq \ell_{max}$: It is readily seen that (\ref{detectioncondition})
implies $F_{r,I}(I) \geq \frac{2 \log n +b_n \sqrt{\log n}}{n}$
for $n$ large enough, hence $\Pr(|F_n(I)-F_{r,I}(I)|\leq 
|\frac{2\log n}{n} -F_{r,I}(I)|) \geq 1-\frac{3}{b_n}$ by Chebyshev.
This implies that with probability converging to 1 we can now
guarantee firstly that $F_n(I) \geq \frac{2\log n}{n}$ and hence
$\ell \leq \ell_{max}$, and secondly, $F_n(I) \leq 2F_{r,I}(I)$, hence
$F_n(I \triangle \tilde{I}) \leq \frac{F_{r,I}(I)}{\sqrt{\ell}}$.
Note that $\tilde{I}$ is a random interval since $\Japp$ is
constructed w.r.t. $F_n$. Hence the above proof for fixed $\tilde{I}$
does not go through any more, but the claim can be established as in
the proof for $A_n^{cond}$ below. There we consider $\tilde{I} \subset I$, 
which can be enforced above while still guaranteeing $ \ell \leq \ell_{max}$.
Alternatively, (\ref{19A}) can be readily extended to cover the case 
$\tilde{I} \not\subset I$. The approximating set $\Iapp$ used for $A_n^{cond}$
differs from $\Japp$ used for $P_n$ in the spacing parameter $d_{\ell}$,
but that is not relevant for the part of the proof below that establishes
$\sqrt{2\lLR (F_0(\tilde{I}),F_n(\tilde{I}))} \geq
\sqrt{2 \log \frac{e}{F_{r,I}(I)}}+B_n$. 

To prove the claim for $A_n^{cond}$ we consider the collection of all
intervals in the approximating set whose endpoints are close
to those of $I$:
$$
{\cal A}(I):= \Bl\{ \tilde{I} \in \Iapp (\ell):\; \tilde{I} \subset I
\;\text{ and } F_n(\tilde{I}) \geq F_n(I) (1-\eta_n/2)\Br\}
$$
where $\eta_n:= \min \Bl(1,\frac{b_n}{2\sqrt{\log e/F_n(I)}} \Br)$ and
$\ell :=\lfloor \log_2 \frac{1}{F_n(I)(1-\eta_n/4)} \rfloor +1$.
Hence $m_{\ell} <nF_n(I)(1-\eta_n/4) \leq 2m_{\ell}$. 
As above one can show that $\ell \in \{ 2,\ldots,\ell_{max}\}$
with probability converging to 1. As in Lemma~2 of Chan and Walther~(2013)
one finds
\be \label{17.1}
\frac{ \# {\cal A}(I)}{\# \Iapp} \; \geq \; C \frac{\eta_n^2 F_n(I)}{
\Bl( \log_2 e/F_n(I) \Br)^{8/5}}
\ee
Standard considerations using Lemma~\ref{approx} and (\ref{19A}) show that
the event $\Bl\{\inf_{\tilde{I} \in {\cal A}(I)} 1\Bl(F_n(\tilde{I})>
F_0(\tilde{I})\Br)=1 \Br\}$ has probability converging to 1, hence on this 
event
\begin{align*}
\inf_{\tilde{I} \in {\cal A}(I)} & \sqrt{2 \lLR \Bl(F_0(\tilde{I}),
F_n(\tilde{I}) \Br)} \;\geq \; \inf_{\tilde{I} \in {\cal A}(I)}
\sqrt{n}\frac{F_n(\tilde{I})-F_0(\tilde{I})}{\sqrt{F_0(\tilde{I}) \vee
F_n(\tilde{I})}} \;\text{ by (\ref{Tay})}\\
& \geq \inf_{\tilde{I} \in {\cal A}(I)} \sqrt{n}\frac{F_{r,I}(\tilde{I})-
F_0(\tilde{I})}{\sqrt{F_{r,I}(\tilde{I}) \vee F_n(\tilde{I})}}
-\sup_{\tilde{I} \in {\cal A}(I)} \sqrt{n}\frac{F_n(\tilde{I})-
F_{r,I}(\tilde{I})}{\sqrt{F_{r,I}(\tilde{I}) \vee F_n(\tilde{I})}}\\
& \geq \Bigg(\inf_{\tilde{I} \in {\cal A}(I)} \sqrt{n}\frac{F_{r,I}(\tilde{I})
-F_0(\tilde{I})}{\sqrt{F_{r,I}(\tilde{I}) }} \Bigg)
\Bigg(1-O_p\Bl(\frac{1}{\sqrt{\log n}} \Br)\Bigg)
 -O_p(1) \;\;\text{ by (\ref{19A})}\\
& \geq \sqrt{n}\frac{F_{r,I}(I) -
F_0(I)}{\sqrt{F_{r,I}(I) }} \Bl(1-\eta_n/2 \Br) 
\Bigg(1-O_p\Bl(\frac{1}{\sqrt{\log n}} \Br) \Bigg) -O_p(1) \\
& \geq \sqrt{2 \log \frac{e}{F_{r,I}(I) (1-F_{r,I}(I))}} +B_n
\;\;\text{ where } B_n:=b_n/9+O_p(1)
\end{align*}
and where the second to last inequality follows from Lemma~\ref{approx}  since 
\begin{align*}
1-\frac{F_{r,I}(I \triangle \tilde{I})}{F_{r,I}(I)}
&\;=\; \frac{F_{r,I}(\tilde{I})}{F_{r,I}(I)}\;=\;
 \frac{F_n(\tilde{I})}{F_n(I)} \Bigg(1+O_p\Bl(\frac{1}{\sqrt{\log n}} \Br) 
\Bigg) \;\;\text{ by (\ref{19A})} \\
& \geq \Bl(1-\eta_n/2 \Br) \Bigg(1+O_p\Bl(\frac{1}{\sqrt{\log n}} \Br) 
\Bigg)^2 \;\;\text{ by the definition of }{\cal A}(I). 
\end{align*}

Hence $\inf_{\tilde{I} \in {\cal A}(I)} LR_n\Bl( F_0(\tilde{I}),F_n
(\tilde{I}) \Br) \geq \frac{1}{F_{r,I}(I) (1-F_{r,I}(I))}
\exp \Bl\{B_n\Bl(B_n/2 +\sqrt{2\log \frac{e}{F_{r,I}(I)}}\Br)\Br\}$
and so $A_n^{cond} \stackrel{P}{\ra} \infty$ as in the proof of Theorem~3
in Chan and Walther~(2013), using (\ref{19A}).
Since the critical value of $A_n^{cond}$ stays bounded by Proposition~\ref{An}, 
the claim follows. 

It remains to show
\be \label{19A}
\sup_{\tilde{I} \in {\cal A}(I)} \Bigg| 1-\frac{F_{r,I}(\tilde{I})}{
F_n(\tilde{I})} \Bigg|\;=\;O_p\Bl(\frac{1}{\sqrt{\log n}} \Br)
\ee

Denote by $X_{(a)}$ the smallest and by $X_{(b)}$ the largest observation
in $I$. Writing $d:=b-a$ and $U_i=F_{r,I}(X_i)$:
$$
\sup_{\tilde{I} \in {\cal A}(I)} \sqrt{n} \frac{\Bl| F_n(\tilde{I})-
F_{r,I}(\tilde{I}) \Br|}{\sqrt{F_n(\tilde{I})}} 
\; \leq \;2 \max_{j=a,\ldots,a+d} \sqrt{n} \frac{\Bl| U_{(j)}-U_{(a)} -
\frac{j-a}{n} \Br|}{\sqrt{\frac{d}{2n}}} \; =\;O_p(1)
$$
by well known facts. Together with $F_n(\tilde{I}) \geq \frac{\log n}{n}$
for $\tilde{I} \in \Iapp$, this implies (\ref{19A}).
$\hfill \Box$

\subsection*{References}

\begin{description}
\item[]{\sc Arias-Casto, E., Donoho, D.L., and Huo, X.} (2005).
Near-optimal detection of geometric objects by fast multiscale
methods. \ {\sl IEEE Trans. Inform. Th.\ \textbf{51}}, 2402--2425.
\item[]{\sc Chan, H.P.} (2009). Detection of spatial clustering with
average likelihood ratio test statistics. \ {\sl Ann. Statist.\ 
\textbf{37}}, 3985--4010.
\item[]{\sc Chan, H.P. and Walther, G.} (2013). 
Detection with the scan and the average likelihood ratio.
{\sl Statistica Sinica \ \textbf{23}}, 409--428.
\item[]{\sc D\"{u}mbgen, L.} (1998).
New goodness-of-fit tests and their application to nonparametric
confidence sets.
{\sl Ann.\ Statist.\ \textbf{26}}, 288--314.
\item[]{\sc D\"{u}mbgen, L. and Spokoiny, V.G.} (2001).
        Multiscale testing of qualitative hypotheses. \
        {\sl Ann.\ Statist.\ \textbf{29}}, 124--152.
\item[]{\sc D\"{u}mbgen, L. and Walther, G.} (2008).
 Multiscale inference about a density. \
{\sl Ann.\ Statist. \textbf{36}}, 1758--1785.
\item[]{\sc  Glaz, J. and Balakrishnan, N. (ed.)} (1999).
{\sl Scan statistics and Applications.} \
Birkh\"{a}user, Boston
\item[]{\sc Good, I.J. and Gaskins, R.A.} (1980).
Density estimation and bump hunting by the penalized
maximum likelihood method exemplified by scattering and meteorite data
        (with discussion).
{\sl J.\ Amer.\ Statist.\ Assoc.\ \textbf{75}}, 42--73.
\item[]{\sc Hartigan, J.A. and Hartigan, P.M.} (1985).
The DIP test of multimodality.
{\sl Ann.\ Statist.\ \textbf{13}}, 70--84.
\item[]{\sc Hoeffding, W.} (1963). Probability inequalities
for sums of bounded random variables.
{\sl J.\ Amer.\ Statist.\ Assoc.\ \textbf{58}}, 13--30.
\item[]{\sc Kulldorff, M.} (1997). A spatial scan statistic. \
{\sl Commun. Statist. - Theory Meth. \textbf{26}}, 1481--1496.
\item[]{\sc Loader, C.R} (1991).
Large-deviation approximations to the distribution of scan
statistics.
{\sl Adv. Appl. Prob. \textbf{23}}, 751--771.
\item[]{\sc Minnotte, M.C. and Scott, D.W.} (1993).
 The mode tree: a tool for visualization of nonparametric density
features.
{\sl J.\ Comp.\ Graph.\ Statist.\ \textbf{2}}, 51--68.
\item[]{\sc M\"{u}ller, D.W. and Sawitzki, G.} (1991).
Excess mass estimates and tests for multimodality.
{\sl J.\ Amer.\ Statist.\ Assoc.\ \textbf{86}}, 738--746.
\item[]{\sc Neill, D. and Moore, A.} (2004).
A fast multi-resolution method for detection of significant spatial
disease clusters.
{\sl Adv. Neur. Info. Proc. Sys. \textbf{10}}, 651--658.
\item[]{\sc Neill, D.B.} (2009a). An empirical comparison of spatial
scan statistics for outbreak detection.\ {\sl Internat. Journal of
Health Geographics \textbf{8}}, 1--16.
\item[]{\sc Neill, D.B.} (2009b). Expectation-based scan statistics
for monitoring spatial time series data.\ {\sl Internat. Journal of
Forecasting \textbf{25}}, 498--517.
\item[]{\sc Polonik, W.} (1995).
Measuring mass concentrations and estimating density contour clusters -
An excess mass approach.
{\sl Ann.\ Statist.\ \textbf{23}}, 855--881.
\item[]{\sc Rufibach, K. and Walther, G.} (2010). The block criterion 
for multiscale inference about a density, with applications to other
a density, with applications to other multiscale problems. \
{\sl Journal of Computational and Graphical Statistics \textbf{19}}, 
175--190.
\item[]{\sc Shorack, G.R. and Wellner, J.A.} (1986).
{\sl Empirical Processes with Applications to Statistics.}
        Wiley, New York
{\sl Ann.\ Statist. \textbf{23}}, 255--271.
\item[]{\sc van der Vaart, A.W. and Wellner, J.A.} (1996).
{\sl Weak Convergence and Empirical Processes with Applications
to Statistics.}
Springer, New York
\item[]{\sc Walther, G.} (2010). Optimal and fast detection of
spatial clusters with scan statistics. \ 
{\sl Ann.\ Statist.\ \textbf{38}}, 1010-1033.
\end{description}

\bigskip

Correspoing author:\\
Guenther Walther\\
390 Serra Mall\\
Stanford, CA 94040\\
walther@stat.stanford.edu

\end{document}